\documentclass[%
 aps,
 amsmath,amssymb,
 preprint, prb %apl%
]{revtex4-1}
\usepackage{graphicx}
\usepackage{xcolor}
\draft % marks overfull lines with a black rule on the right
\usepackage{array}
\usepackage{placeins} 
\begin{document}

% Use the \preprint command to place your local institutional report number 
% on the title page in preprint mode.
% Multiple \preprint commands are allowed.
%\preprint{}

\title{Spin-to-charge conversion by spin pumping in sputtered polycrystalline  Bi$_x$Se$_{1-x}$} %Title of paper

% repeat the \author .. \affiliation  etc. as needed
% \email, \thanks, \homepage, \altaffiliation all apply to the current author.
% Explanatory text should go in the []'s, 
% actual e-mail address or url should go in the {}'s for \email and \homepage.
% Please use the appropriate macro for the type of information

% \affiliation command applies to all authors since the last \affiliation command. 
% The \affiliation command should follow the other information.

\author{Isabel C Arango\footnote{Equal contribution}}
\affiliation{CIC nanoGUNE BRTA, 20018 Donostia-San Sebastian, Basque Country, Spain.}

\author{Alberto Anadón$^{a,b}$}
\affiliation{Institut Jean Lamour, Université de Lorraine CNRS UMR 7198, Nancy, France}
%\email[alberto.anadon@univ-lorraine.fr]

\author{Silvestre Novoa}
\affiliation{Institut Jean Lamour, Université de Lorraine CNRS UMR 7198, Nancy, France}

\author{Van Tuong Pham}
\affiliation{IMEC, Kapeldreef 75, Leuven, Belgium B-3001}

\author{Won Young Choi}
\affiliation{CIC nanoGUNE BRTA, 20018 Donostia-San Sebastian, Basque Country, Spain.}
\affiliation{VanaM Inc, 21-1 Doshin-ro 4-gil, Yeongdeungpo-gu, Seoul, Korea.}

\author{Junior Alegre}
\affiliation{Institut Jean Lamour, Université de Lorraine CNRS UMR 7198, Nancy, France}
\affiliation{Facultad de Ciencias, Universidad Nacional de Ingeniería, Rímac, 15333 Peru}

\author{Laurent Badie}
\affiliation{Institut Jean Lamour, Université de Lorraine CNRS UMR 7198, Nancy, France}

\author{Andrey Chuvilin}
\affiliation{CIC nanoGUNE BRTA, 20018 Donostia-San Sebastian, Basque Country, Spain.}
\affiliation{IKERBASQUE, Basque Foundation for Science, 48009 Bilbao, Basque Country, Spain.}

\author{Sébastien Petit-Watelot}
\affiliation{Institut Jean Lamour, Université de Lorraine CNRS UMR 7198, Nancy, France}

\author{Luis E Hueso}
\affiliation{CIC nanoGUNE BRTA, 20018 Donostia-San Sebastian, Basque Country, Spain.}
\affiliation{IKERBASQUE, Basque Foundation for Science, 48009 Bilbao, Basque Country, Spain.}

\author{Fèlix Casanova\footnote{alberto.anadon@univ-lorraine.fr, juan-carlos.rojas-sanchez@univ-lorraine.fr, f.casanova@nanogune.eu}}
\affiliation{CIC nanoGUNE BRTA, 20018 Donostia-San Sebastian, Basque Country, Spain.}
\affiliation{IKERBASQUE, Basque Foundation for Science, 48009 Bilbao, Basque Country, Spain.}

\author{Juan-Carlos Rojas-Sánchez$^{b}$}
\affiliation{Institut Jean Lamour, Université de Lorraine CNRS UMR 7198, Nancy, France}

%\email[juan-carlos.rojas-sanchez@univ-lorraine.fr,f.casanova@nanogune.eu]

%(Sin orden concreto tenemos a: Seb?, Junior, Alberto, Carlos, Isabel, Félix, Van-Tuong PHAM, nos dejamos a alguien? )

%\email[]{Your e-mail address}
%\homepage[]{Your web page}
%\thanks{}
%\altaffiliation{}

% Collaboration name, if desired (requires use of superscriptaddress option in \documentclass). 
% \noaffiliation is required (may also be used with the \author command).
%\collaboration{}
%\noaffiliation

\date{\today}

\begin{abstract}
% insert abstract here

Topological materials are of high interest due to the promise to obtain low power and fast memory devices based on efficient spin-orbit torque switching or spin-orbit magnetic state read-out. In particular, sputtered polycrystalline Bi$_x$Se$_{1-x}$ is one of the materials with more potential for this purpose since it is relatively easy to fabricate and has been reported to have a very high spin Hall angle. We study the spin-to-charge conversion in Bi$_x$Se$_{1-x}$ using the spin pumping technique coming from the ferromagnetic resonance in a contiguous permalloy thin film. We put a special emphasis on the interfacial properties of the system. Our results show that the spin Hall angle of Bi$_x$Se$_{1-x}$ has the same sign as  the one of Pt. The charge current arising from the spin-to-charge conversion is, in contrast, lower than Pt by more than one order of magnitude. We ascribe this to the interdiffusion of Bi$_x$Se$_{1-x}$ and permalloy and the changes in chemical composition produced by this effect, which is an intrinsic characteristic of the system and is not considered in many other studies.

\end{abstract}

\pacs{Spintronics, Spin Hall effect, Topological materials}% insert suggested PACS numbers in braces on next line

\maketitle %\maketitle must follow title, authors, abstract and \pacs

% Body of paper goes here. Use proper sectioning commands. 
% References should be done using the \cite, \ref, and \label commands
\section{Introduction}
\label{sec:introduction}

Spintronics is a promising beyond-CMOS technology that exploits the spin degree of freedom of the electron in the form of spin currents.\cite{maekawa2017spin} Information stored in magnetic materials can be transferred and manipulated by mastering these spin currents, leading to the realization of devices such as magnetic random-access memory (MRAM) \cite{bhatti2017spintronics} and to proposals that would integrate magnetic memory and logic operations\cite{manipatruni2019scalable,vaz2021functional,pham2020spin}. A very convenient way to generate spin currents is by exploiting the spin-orbit coupling (SOC) present in a variety of systems that leads to  charge-to-spin current conversion, for example, the spin Hall effect (SHE) in bulk materials \cite{sinova2015spin,anadon2022thermal} or the Edelstein effect (EE) in Rashba interfaces and topologically protected surface states.\cite{rojas2019compared,bihlmayer2022rashba,zhang2016conversion,anadon2021engineering,anadon2020spin} Reciprocally, spin currents can be detected with the inverse effects that lead to spin-to-charge current conversion. Therefore, the search for systems with more efficient spin-charge interconversion is crucial for different technologies, from the new generation of MRAMs that exploits the SHE and/or EE to switch the magnetic element\cite{manchon2019current, shao2021roadmap} to the spin-based logic that uses the inverse effects to read-out the magnetic element.\cite{pham2020spin} The efficiency of the spin-to-charge conversion in the case of SHE and its inverse (ISHE) is known as the spin Hall angle ($\theta$\textsubscript{SH}) and for the inverse Edelstein effect (IEE) is the inverse Edelstein length ($\lambda$\textsubscript{IEE}). These efficiencies are defined as the ratio between the charge current and the spin current, which are dimensionless in 3D but have units of length in 2D.\cite{rojas2019compared} 

In this regard, topological insulator materials such as Bi$_2$Se$_3$ have drawn much attention due to their unique properties. In particular, the spin-momentum locking at the topologically protected surface states makes them desirable for spin-charge interconversion in spintronics devices.\cite{manipatruni2019scalable,deorani2014observation,wang2017room,mellnik2014spin,jamali2015giant} Even though these topological properties are supposed to be linked to an epitaxial growth and structure,\cite{moore2010birth} some works report large spin-charge interconversion in sputtered polycrystalline Bi$_x$Se$_{1-x}$ (BiSe).\cite{dc2019observation,dc2018room,dc2019room} According to Mahendra DC et al.\cite{dc2018room}, the granular structure possessed by the sputtered BiSe layers present quantum confinement and thus a high efficiency, although quantum phenomena are challenging to be evidenced at room temperature.\cite{doi:10.1126/science.1137201} 

BiSe would be a promising candidate to be placed in the magnetic state read-out node of the MESO (Magneto Electric Spin-Orbit) logic device due to its high resistivity and large spin-charge interconversion.\cite{manipatruni2018beyond} The conversion efficiency of this material has been estimated using different techniques, such as spin pumping,\cite{dc2019observation,dc2019room} DC planar Hall,\cite{dc2018room} spin-torque ferromagnetic resonance,\cite{dc2018room} spin-orbit torque (SOT) current-induced magnetic switching\cite{dc2018room} and second harmonic Hall measurements.\cite{dc2018room} However, in all these approaches, BiSe needs to be in contact with a ferromagnet (FM) and it is extremely challenging to obtain clean interfaces in a BiSe/FM stack.  When this FM is metallic, a large intermixing at the interface between these materials is a frequent phenomenon.\cite{bonell2020control,walsh2017interface,choi2022all} Such intermixing at the interface affects the material characterization because spin currents are pumped/injected through an additional layer, leading to a poor estimation of the relevant spin transport parameters.\cite{choi2022all}  This is the case not only for sputtered films but also when growing BiSe by techniques such as molecular beam epitaxy in ultra-high vacuum as observed in the growth of BiSe onto insulating ferrimagnets such as yttrium iron garnet, where even though an atomically ordered BiSe layer is obtained with a thickness of a few monolayers, an amorphous layer of about 1 nm at the interface between them has been observed by several groups.\cite{PhysRevLett.117.076601,fanchiang2018strongly} This low-quality interface in YIG/BiSe leads to a low conversion efficiency with a $\lambda$\textsubscript{IEE} of 0.1 nm, one order of magnitude smaller than for other topological insulators (TIs) such as $\alpha$-Sn.\cite{PhysRevLett.116.096602} Besides, theoretical predictions suggest that if this TI is in direct contact with a metallic ferromagnet a hybridization is produced destroying the helical spin texture or spin-momentum locking.\cite{PhysRevB.94.014435} 

%\textcolor{orange}{La referencia \cite{dc2019observation} usa Spin Pumping para determinar $\lambda$\textsubscript{IEE} en Sputtered Bismuth Selenide Thin Films at Room Temperature} 

Here, we study the spin-to-charge current conversion by spin pumping ferromagnetic resonance (SP-FMR) in sputtered Bi$_x$Se$_{1-x}$/permalloy (Py) bilayers and its opposite stacking order. Our results show that $\theta$\textsubscript{SH} of Bi$_x$Se$_{1-x}$ has the same sign to that of Pt, in contrast with a recent study by Mendes et al.,\cite{mendes2021unveiling} and it is lower than in other reports by more than one order of magnitude.\cite{dc2018room,choi2022all} Structural characterization of the samples by transmission electron microscopy (TEM) performed at the interfaces of the bilayers helps us to understand why sputtered films show this low spin conversion and how the interface and the different stoichiometries of the films could contribute to a drastic overestimation of the spin-charge interconversion efficiency.  

\section{Methods}
\label{sec:methods}
\subsection{Sample growth}

All samples were grown on Si/SiO$_2$(300 nm) substrates by sputtering deposition at room temperature. Targets of Bi$_2$Se$_3$ (99.999\% pure) and Py (Fe$_{20}$Ni$_{80}$, 99.95\% pure) were used in an ultrahigh vacuum seven-target AJA sputtering system with a base pressure of $3\cdot 10^{-8}$ Torr. Bi$_2$Se$_3$ was radiofrequency (RF) sputtered at a 35 W power and a 3 mTorr Ar pressure to yield a deposition rate of 0.09 \AA /s. The Py layers were sputtered at 100 W DC power and a 3 mTorr Ar pressure to yield a deposition rate of 0.08 \AA /s. The bilayers were capped with 5 nm of Al$_2$O$_3$ (200 W RF at 3 mTorr Ar pressure). The bilayers, including the capping, were grown \emph{in situ}. Sample stacks are always written in this document from left to right corresponding from bottom to top; i.e. BiSe/Py corresponds to the BiSe being grown on top of the substrate and Py on top of BiSe.

\subsection{Device fabrication and SP-FMR measurements}
\begin{figure}
\includegraphics[width=0.75\textwidth]{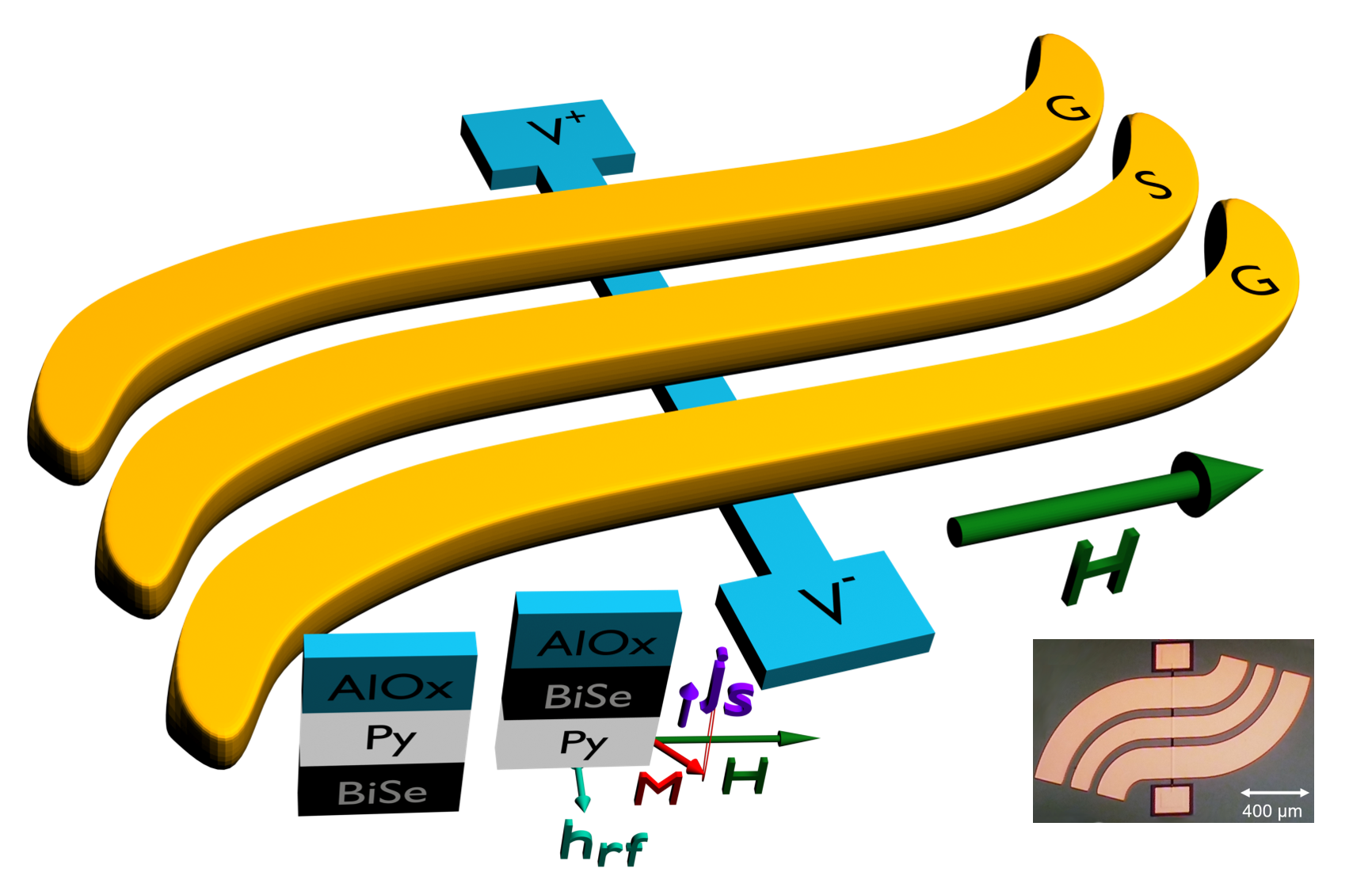}
\caption{\label{fig:SPschematics}\textbf{Spin Pumping device.} A DC field (\textit{H}) is applied in-plane perpendicular to the sample while an RF electric current is injected in the coplanar waveguide (yellow) producing an RF magnetic field (\textit{h}\textsubscript{RF}) that produces the precession of the magnetization ($M$) in the ferromagnetic layer of the sample (Py). This creates a spin current in the out of plane direction of the sample (\textit{j}\textsubscript{S}) that is injected in the material with SOC (BiSe) and converted into a voltage ($V$\textsubscript{SP}=$V^+-V^-$) by ISHE or IEE. Inset: Optical image of a similar device that the ones used in this study. The scalebar represents 400 $\mu$m.}
\end{figure}

The spin pumping devices were prepared using conventional UV lithography. The full stack was first patterned and subsequently ion milled controlling the milled thickness by an ion mass spectrometer using a 4-wave IBE14L01-FA system. After that, in a second step, an insulating SiO$_2$ layer with a thickness of 200 nm was grown by RF sputtering using a Si target and Ar$^+$ and O$^{2-}$ plasma in a Kenositec KS400HR PVD. In a third lithography step, the contacts were patterned and evaporated using an evaporator PLASSYS MEB400S. The dimensions of the active bar (see the blue part in figure \ref{fig:SPschematics}) are 10 $\times$ 600 $\mu$m. Due to the small width of the bar, we do not expect significant artifacts from rectification effects in the spin pumping signal.\cite{sanchez2013spin,PhysRevLett.116.096602,martin2022suppression} The geometry of the devices, including the thickness of the insulating SiO$_2$, the dimensions of the coplanar waveguide and the lateral dimensions of the milled samples are similar in all the devices shown in this study to reliably compare the SP voltage. Since the sample stack extends beyond the CPW, the driving RF field is not homogeneous along the magnetic wire. However, this inohomegeity does not affect to the shape of the measured spin pumping signal and the main findings of this manuscript, since it is similar for all the devices.

The spin pumping measurements were performed using a probe station with in-plane DC magnetic field (\textit{H}) up to 0.6 T provided by an electromagnet. A sketch of the SP-FMR device is shown in figure \ref{fig:SPschematics}. In this system, an RF current with a fixed frequency (\textit{f}) of the order of GHz is injected into the coplanar waveguide generating an RF magnetic field on the sample (\textit{h}\textsubscript{RF}). At certain combinations of the DC field and the frequency of \textit{h}\textsubscript{RF} this field drives the ferromagnetic resonance in the Py, and by the spin-pumping effect,\cite{PhysRevLett.112.106602,PhysRevB.81.094409,PhysRevLett.88.117601,RevModPhys.77.1375,PhysRevLett.110.217602} the precession of the magnetization produces a transverse spin current that is injected from the Py into the non-magnetic layer (i.e. BiSe or Pt). This spin current is then converted in the non-magnetic material into a charge accumulation by means of the inverse SHE (ISHE) or the inverse EE (IEE). We can measure this voltage (\textit{V}\textsubscript{SP}) by modulating the RF power injected in the coplanar waveguide and using a lock-in voltmeter that is matched to this modulation while sweeping the external \textit{H}. We use a power modulation with a sine function, where the depth was 100\% and the modulation frequency was 433 Hz. When the system reaches the resonance condition, the measured voltage exhibits a characteristic Lorentzian curve symmetric around the resonance field (\textit{H}\textsubscript{res}) (see figure \ref{fig:SPexperimental}). The voltage signal from SP-FMR always shows in the real part of the lock-in. We always monitor both the real and imaginary parts of the voltage and never change the phase of the measurement. Any transport effects are fast enough to appear without delay in the measurement, while any other thermal effects that might be slower would show in the imaginary part. 

To obtain the effective magnetization (\textit{M}\textsubscript{eff}) and the Gilbert damping ($\alpha$) of the ferromagnetic layer, we analyze the position of the SP-FMR resonance by observing the peak in \textit{V}\textsubscript{SP}. We analyze the center (\textit{H}\textsubscript{res}) and width ($\Delta$H) of this peak as a function of frequency using the conventional method of fitting the voltage to a sum of a symmetric and an antisymmetric Lorentzian functions:

\begin{equation}
\begin{split}
    V_{SP}=V_{offset}+V_{sym}\frac{\Delta H^2}{\Delta H^2+(H-H_{res})^2}+ \\ V_{antisym}\frac{\Delta H(H-H_{res})}{\Delta H^2+(H-H_{res})^2}
\end{split}
\end{equation}

The antisymmetric part of the signal is negligible in our measurements and we consider only the symmetric part in the fit. Then, we consider the Kittel formula for an in-plane easy axis.

\begin{equation}
    f=\frac{\gamma}{2\pi}\sqrt{(H_{res}+H_{uni})(H_{res}+H_{uni}+M_{eff})},
    \label{eq:Meff}
\end{equation}
where $\gamma$ is the gyromagnetic ratio and \textit{H}\textsubscript{uni} is a small in-plane uniaxial magnetic anisotropy. The damping is obtained considering the linear dependence of $\Delta$H with the frequency as 
\begin{equation}
\Delta H=\Delta H_0+\alpha\frac{2\pi f}{\gamma} 
\end{equation}

Here,  $\Delta$\textit{H}\textsubscript{0} is the frequency-independent inhomogeneous contribution. Comparing $\alpha$ in our system with a reference Py thin film, we can estimate the spin transparency of the interface between Py and the spin conversion layer. This is given by the real part of the effective spin-mixing conductance, $g_{\uparrow\downarrow}$:
\begin{equation}
   g_{\uparrow\downarrow} = \frac{4\pi M_{s}t_{Py}}{g\mu_B}\left(\alpha_{Py/BiSe}-\alpha_{Py}\right),
\end{equation}
where \textit{M}\textsubscript{s} is the saturation magnetization, \textit{t}\textsubscript{Py} is the thickness of the Py layer, \textit{g} is the Landé factor, and $\mu_B$ is the Bohr magneton. $\alpha$\textsubscript{Py/BiSe} and $\alpha$\textsubscript{Py} correspond to the damping of the Py/BiSe bilayer and the reference Py thin film ($\alpha$\textsubscript{Py}=0.0073(2)).

\subsection{Structural characterization by TEM}
Cross-sectional samples for the transmission electron microscopy energy-dispersive X-ray spectroscopy ((S)TEM-EDX) analysis were prepared by a standard focused ion beam (FIB) lamellae preparation method: the surface of the deposited samples was protected first by e-beam followed by i-beam Pt deposition, the lamellae was cut and lifted out onto a Mo 3-post half-grids. Mo grids were selected to avoid an overlap of Ni \textit{K}\textsubscript{$\beta$} line (Ni is one of the elements of interest) with Cu \textit{K}\textsubscript{$\alpha$} line, which is a typical artifact in EDX spectra, if a sample is held on a Cu grid. The cross-sections were studied on a Titan 60-300 TEM (FEI, Netherlands) at 300 kV in STEM mode. EDX spectral images were acquired using EDAX RTEM spectrometer. Element distribution maps were obtained by  multiple linear least-squares (MLLS) deconvolution of spectral images utilizing simulated spectral components. 

%In this antenna we injected a GHz rf current which generates the radio-frequency field hrf on the sample. A static magnetic field ranging from 0 to 0.5 T can also be applied in plane. Depending on the value of the applied field, hrf could excite the magnetization at resonance. Through the spin-pumping effect, the precession of the magnetization yield to the creation of a pure transverse spin current js that is injected from the CoFeB layer into the Ir or Pt layer.
\section{Results}
\label{sec:results}

\begin{figure*}
\includegraphics[width=\textwidth]{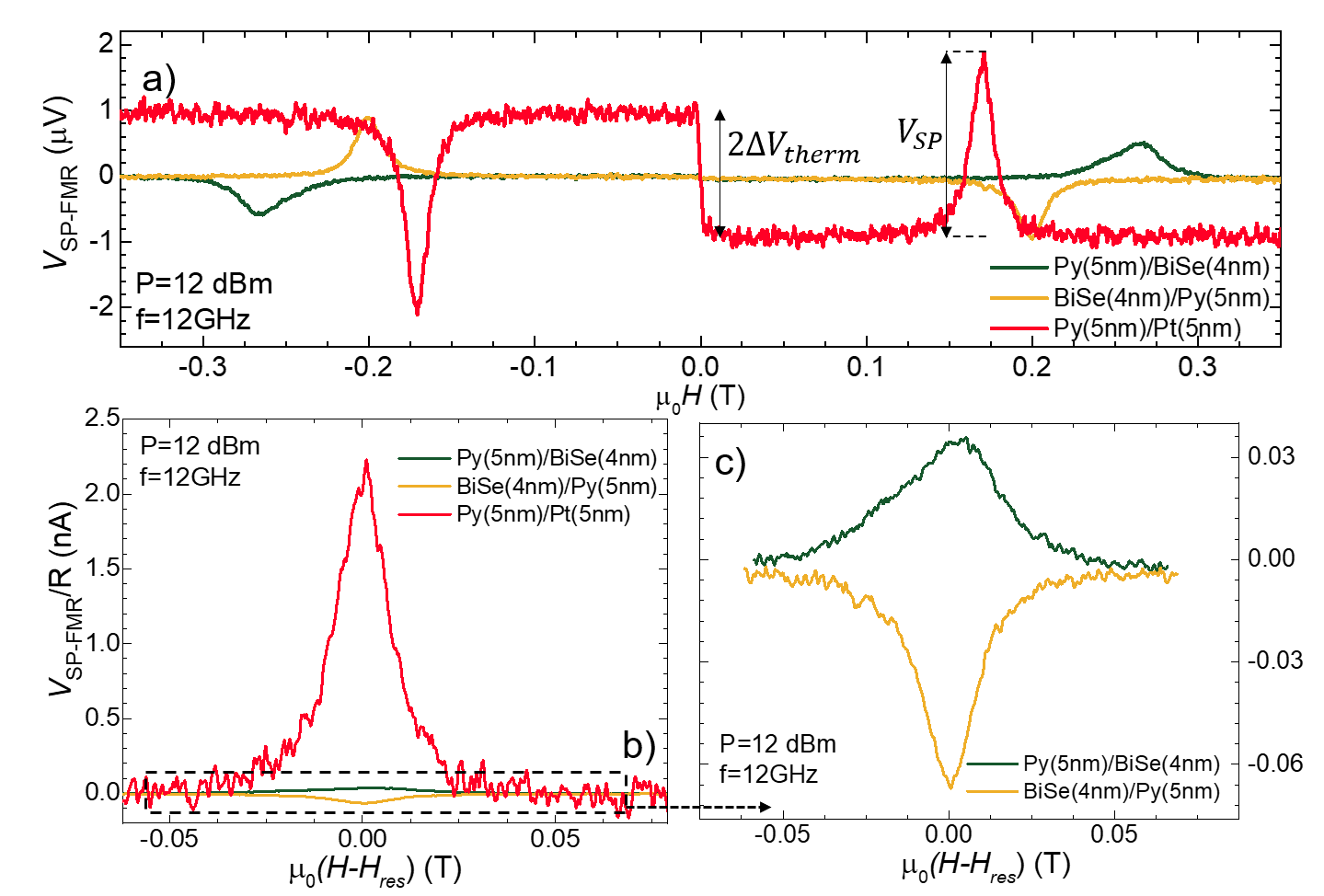}
\caption{\label{fig:SPexperimental} \textbf{Spin pumping voltages in Py/BiSe, BiSe/Py and Py/Pt.} a) Spin pumping voltage as a function of the applied field for the Py/BiSe, BiSe/Py and a reference Py/Pt stacks at 12 GHz and a power of 12 dBm. In addition to the spin pumping voltage (\textit{V}\textsubscript{SP}) at the ferromagnetic resonance condition, we can also see a jump around $H=0$ T due to thermovoltage (\textit{V}\textsubscript{therm}). This thermal voltage is much smaller in the case of the BiSe layers compared to the Py/Pt sample. b) Comparison of the spin pumping voltages of the three bilayers normalized by the sample resistance. The sign of the spin Hall angle in the Py/BiSe system is the same as the one in Py/Pt. c) Detail of the Py/BiSe and BiSe/Py spin pumping voltages}
\end{figure*}

We performed SP-FMR measurements in the BiSe/Py and Py/BiSe samples and in a reference Py/Pt sample, shown in figure \ref{fig:SPexperimental}. In these measurements, thermal and other artifacts can be relevant and need to be accounted for.\cite{thermalIguchi} In figure \ref{fig:SPexperimental} a) we observe a voltage change of 2$\Delta$\textit{V}\textsubscript{therm} at $H=0$, which corresponds to the contribution from the anomalous Nernst effect (ANE) and the spin Seebeck effect (SSE) from the bilayer. In the case of Py/Pt, this jump is slightly smaller than the SP peak, while in the BiSe stack, it is much smaller. Considering that the precession angle in Py is typically a few degrees\cite{PhysRevB.85.014410}, we can safely assume that the thermovoltages are not significant in resonance conditions from this assumption. Asymmetries in the placement of the CPW structure with respect to the bar in which the sample stack is patterned could give rise to other thermal contributions to the voltage. However, since we observe a clear Lorentzian line shape in the SP voltage and a clear sign change with the same amplitude with for negative and positive applied magnetic fields, only Nernst, ANE or SSE induced thermovoltages caused by a change in the temperature gradient profile due to absorption of the RF power by the Py could arise.\cite{thermalIguchi} Any changes in the temperature profile due to thermal transport by spinwaves can be neglected due to the thin Py layer.\cite{thermalIguchi}  Figure \ref{fig:SPexperimental} b) shows the SP-FMR voltage divided by the sample two-probe resistance, i.e. the current coming from the spin-to-charge conversion in BiSe. The sign of the voltage is in agreement with a previous study,\cite{mendes2021unveiling} but the opposite to others.\cite{dc2019observation,dc2019room} These discrepancies in reproducibility can have an interfacial origin and thus a careful structural analysis of the samples is needed. The magnitude of \textit{V}\textsubscript{SP/R} is smaller than that of a reference Py/Pt layer as shown in figure \ref{fig:SPexperimental}. In contrast, previous results reported a very highly efficient spin-to-charge conversion, with a $\theta$\textsubscript{SH} about 200 times larger than the one of Pt.\cite{dc2018room} Figure \ref{fig:SPexperimental} c) shows that the sign of the SP-FMR current changes when the stack is inverted, as expected in SP-FMR measurements. Remarkably, the height and width of the peak is significantly different for both stacks, which cannot be explained if we consider the interface and sample quality similar for both samples. More specifically, the charge current produced under the same conditions, i.e. 12 dBm (15.85 mW) and 12 GHz, is -1.153$\pm$0.002 nA for the Py/Pt stack, while only 0.040$\pm$0.001 and -0.062$\pm$0.001nA for the Py/BiSe and BiSe/Py, respectively, as shown in table \ref{tab:meff}.

\begin{table}[htb]
\centering
\begin{tabular}{||p{3.50cm}|p{2.25cm}|p{1.5cm}|p{1.5cm}|p{1.8cm}|p{2.3cm}|p{2.3cm}||}
%\begin{tabular}{ | m{6cm} | m{4cm}| m{4cm} | } 
\hline\hline
Sample            & \textit{M}\textsubscript{eff}\newline (emu/cm$^3$) & \textit{H}\textsubscript{uni} (G) & $\Delta$\textit{H}\textsubscript{0} (G) & $\alpha$ & $\textit{g}_{\uparrow\downarrow}$ \newline(m$^{-2}\cdot10^{19}$) & \textit{V}\textsubscript{SP}/\textit{R} (nA)   \\ \hline\hline
Py(5nm)/BiSe(4nm) & 316(3)   &     -60(10)  &   12(3)   & 0.0453(1) & 3.88(1) & 0.040(1)   \\ \hline
BiSe(4nm)/Py(5nm) & 514(1)   &       -13(2)  &   6(1)   & 0.0220(3) & 2.44(1) & -0.062(1)   \\ \hline
Py(5nm)/Pt(5nm) & 701(2)     &     -8(2)  &   3.8(0.1)   & 0.0269(2) & 4.43(1) & 2.228(3)   \\ \hline\hline
Py(6nm)/Au & 628(1)     &     -8(1)  &   5.6(0.4)   & 0.0073(2) & - & -   \\ \hline\hline
\end{tabular}
\caption{Effective magnetization, uniaxial in-plane magnetic anisotropy, frequency-independent inhomogeneous contribution ($\Delta$\textit{H}\textsubscript{0}),  damping, effective spin-mixing conductance and charge current generated by spin-to-charge conversion (at 12 GHz) for Py/BiSe, BiSe/Py, Py/Pt and a Py/Au reference sample. For the estimation of $\textit{g}_{\uparrow\downarrow}$, we are considering that all damping enhancement comes from the spin pumping effect, which is not accurate, as discussed in the main text.}
\label{tab:meff}
\end{table}

\begin{figure*}
\includegraphics[width=\textwidth]{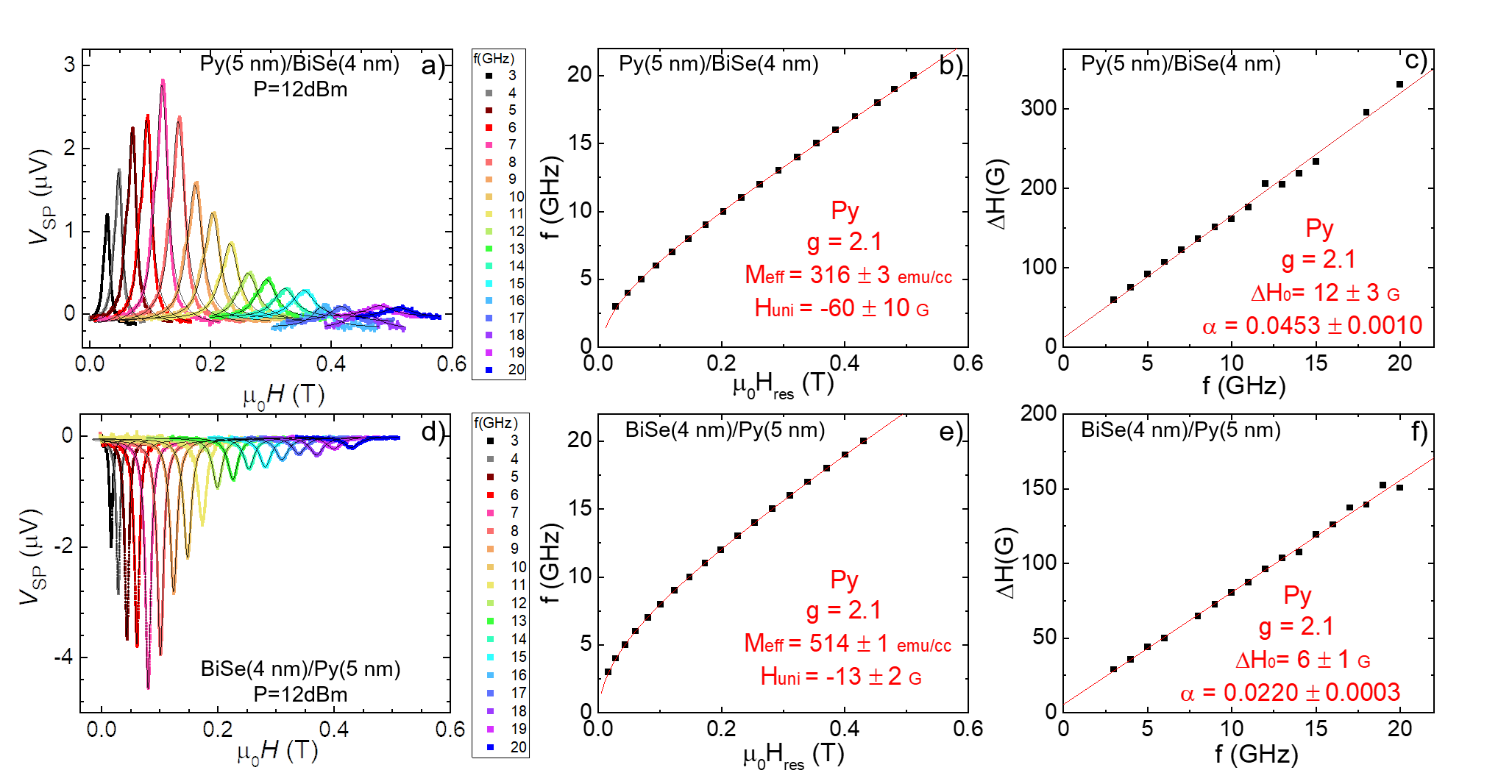}
\caption{\label{fig:SPfreq} \textbf{Evolution of the spin pumping measurements as a function of frequency.} a) evolution of the spin pumping voltage with the frequency of the RF excitation as well as the extracted b) resonance field and c) line width for the Py/BiSe and the d), e), f) BiSe/Py samples respectively.}
\end{figure*}

\begin{figure*}
\includegraphics[width=\textwidth]{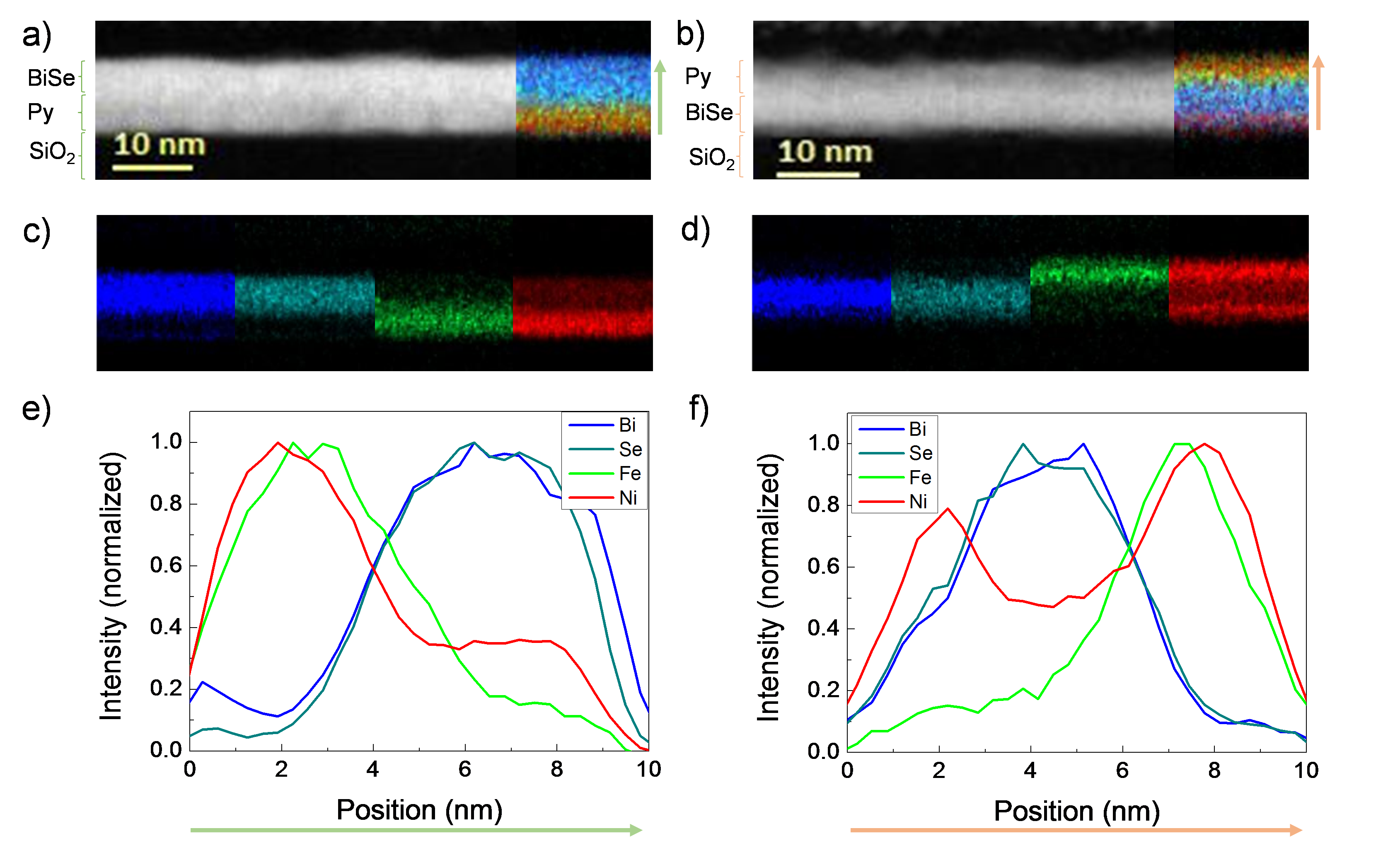}
\caption{\label{fig:TEM} \textbf{Interfacial structure and chemical characterization of thin film bilayers.} a) Py(5nm)/BiSe(4nm) and b) BiSe(4nm)/Py(5nm) by TEM. Elemental EDX maps of c) Py/BiSe and d) BiSe/Py with the essential elements: Bi (blue), Se (dark Cyan), Fe (green) and Ni (red), with the SiO$_2$ substrate at the bottom and the capping layer at the top. Elemental normalized profiles of e) Py/BiSe and f) BiSe/Py, starting from the substrate (left) to the top (right). The green (a)) and orange (b)) arrows indicate the direction of the scan.}
\end{figure*}

In figure \ref{fig:SPfreq}, we show the frequency dependence of the SP-FMR voltage for a fixed RF power of 12 dBm to obtain the values of \textit{M}\textsubscript{eff}, $\alpha$ and $\textit{g}_{\uparrow\downarrow}$, (given in table \ref{tab:meff}) using the fits to Eq. 2-4. We have measured independently \textit{g} of Py for the same range of thickness, yielding a value of 2.10, similarly to the other stacks in this study. We can observe that both the magnetic properties of the sample (\textit{M}\textsubscript{eff}, $\alpha$) and the interfacial spin transport ones ($\textit{g}_{\uparrow\downarrow}$) are significantly different between the two BiSe stacks. We can also observe that \textit{M}\textsubscript{eff} is lower in both BiSe stacks than in the Pt stack, which presents a value closer to the one of bulk Py.\cite{pybulk} The obtained values of $H_{uni}$ are small in all cases, suggesting that the Py layers do not have a significant anisotropy in the film plane. Even though this is true, we observe a larger value for the case of the Py/BiSe sample (-60$\pm$10 G) compared to the BiSe/Py (-13$\pm$2 G).

It has been widely acknowledged that the interface plays a crucial role in the injection of spin currents between different materials.\cite{PhysRevLett.61.2472,bonell2020control,he2021enhancement,soumyanarayanan2016emergent,PhysRevLett.112.106602,zhang2015role,cosset2021evidence} In order to further explore the origin of these  differences, we study the structural properties of our samples by comparing how the interface changes with the stacking order in a cross-section of the sample observed by TEM. We show the two opposite stacking orders Py/BiSe and BiSe/Py in figures \ref{fig:TEM}a) and b), respectively. In both cases, a uniform and continuous material deposition is observed. Additionally, the chemical distribution has been characterized by EDX. Figures \ref{fig:TEM} c) and d) show the different elemental maps obtained by EDX. Figures \ref{fig:TEM}e) and f) display the normalized elemental profiles for each stack. A strong diffusion of Ni through the BiSe layer is observed in both cases, in agreement with previous studies.\cite{walsh2017interface} A clear shift of the Ni (red line) curve is observed in the elemental profiles, being larger when BiSe is at the bottom of Py (fig. \ref{fig:TEM} f)), with Ni penetrating through all the BiSe layer and accumulating at the bottom. As a consequence, Fe is shifted in both stacking orders (green line) creating a Fe-rich interface with the BiSe. The interface between BiSe and Py is completely modified in both stacking orders, and this also alters the spin injection efficiency for spin transport measurements. By comparing our spin pumping and EDX-STEM results, we can observe that the bilayer with the highest interdiffusion, BiSe/Py, also has the highest charge current produced since we have a more complex structure with a large intermixing of chemical elements at the interface. Quantification of the spin-charge interconversion efficiency by assuming a single bilayer would be meaningless.

The migration of Ni in the Py/BiSe sample can also explain the origin of the reduced effective magnetization as well as the change in $\alpha$. The Gilbert damping increases 2-fold for the Py/BiSe and 5-fold in the BiSe/Py samples in comparison to Py/Pt. This could be due to a combination of the interdiffused interface and the different SOC in the BiSe. The different compositions of the Py close to the interface could also have a relevant impact on the spin pumping voltage since it is very sensitive to the transparency of the interface. The migration of the elements in both samples could lead to a gradient of magnetic composition, since the moment per atom of Ni (bulk saturation magnetization \textit{M}\textsubscript{s}=485 emu/cm$^3$) and Fe (bulk \textit{M}\textsubscript{s}=1707 emu/cm$^3$)\cite{kittel1996introduction} are very different. In fact, while $\alpha$ doubles for the Py/BiSe sample compared to the BiSe/Py one, $\Delta$\textit{H}$_0$ is also two times larger, indicating that the origin of this increase is not due to an interfacial effect but is related to a change of properties of the magnetic layer. \textit{M}\textsubscript{eff} is usually different to \textit{M}\textsubscript{s} and changes in the sample anisotropy typically lead to a reduction in \textit{M}\textsubscript{eff}, but not in \textit{M}\textsubscript{s}. In the thin film limit and when the magnetic anisotropy is negligible, \textit{M}\textsubscript{eff} and \textit{M}\textsubscript{s} are similar. In this sense, a gradient of composition in Ni and Fe could also be the origin of an out of plane anisotropy that causes a reduction in \textit{M}\textsubscript{eff} similar to what we observe in both stacks. In contrast, regarding the change in composition, one could expect that a reduction of the Ni percentage in the Py layer would produce an increase in \textit{M}\textsubscript{eff}. Regarding the reduced charge current generated by spin pumping, is relevant to consider the role of the Ni migration and how this could induce spin currents coming from the Ni and Fe inside the BiSe and even an opposing voltage coming from the migrated Ni layer in the case of the BiSe/Py stack. These deviations from the conditions considered for the determination of the parameters shown in table \ref{tab:meff} directly affect the reliability of the obtained values. For example, the model to obtain $g_{\uparrow\downarrow}$ implies that both materials at the interface have the same composition as the magnetic layer and we observe a mixed interface. Additionally, the extraction of \textit{M}\textsubscript{eff} from equation \ref{eq:Meff} implies that the magnetic layer is homogeneous or the voltage measured comes from the spin conversion in BiSe, while in reality, we do not have a homogenous magnetic layer or a pure BiSe film.  Furthermore, given that the magnetic properties of the magnetic layer are different for the different stacks, the spin current generated can also vary significantly even though the \textit{h}\textsubscript{RF} is similar.

%\FloatBarrier
\section{Conclusions}
\label{sec:conclusions}
In summary, we show that the charge current generated by spin pumping  in sputtered Bi$_x$Se$_{1-x}$ has the same sign to the one of Pt and is significantly lower than in a Py/Pt reference sample. By measuring the frequency dependence of the spin pumping voltage, we compare the different magnetic and interfacial properties of a Py/Bi$_x$Se$_{1-x}$ and a Bi$_x$Se$_{1-x}$/Py bilayers and observe that the effective magnetization and Gilbert damping are very different for both stacks, including a small unidirectional anisotropy in the Py in both cases.

We then study the structural and composition of both systems by transmission electron microscopy energy-dispersive X-ray spectroscopy, finding a strong interdiffusion characteristic of Bi$_x$Se$_{1-x}$ thin films. The chemical composition of the magnetic layers and the interfaces are not homogeneous and, therefore, the models used to characterize the properties of the system are not valid anymore.  These inhomogeneities in the interfaces and the films enhance the Gilbert damping constant and reduce the effective magnetization, which would lead to an incorrect estimation of the spin conversion parameters. Additionally, the compositional gradient in the Py layer might induce anisotropies that reduce the effective magnetization in both stacks. 

Our work highlights the importance to study the interfacial and compositional properties of Bi$_x$Se$_{1-x}$ systems for spin conversion since they tend to produce systems with very high interdiffusion and thus the performance is highly dependent on the quality of the layers and interfaces as well as their stoichiometric composition.

\section{Acknowledgements}
\label{sec:acknowledgements}
The work in Institut Jean Lamour was funded by the French National Research Agency (ANR) through project ANR-19-CE24-0016-01 ‘Toptronic ANR’. The work in CIC nanoGUNE BRTA was funded by Intel Corporation under the ‘FEINMAN’ Intel Science Technology Center and by the Spanish MICINN under project No. PID2021-122511OB-I00 and Maria de Maeztu Units of Excellence Programme No. CEX2020-001038-M. Devices in the present study were patterned at Institut Jean Lamour's clean room facilities (MiNaLor). These facilities are partially funded by FEDER and Grand Est region through the RANGE project.

\newpage
\appendix
\counterwithin{figure}{section}

\section{Magnetometry}
\FloatBarrier

Here we add magnetometry data in a new set of samples grown under the same conditions and measured by vibrating sample magnetometry (VSM).

\begin{figure}[htb]
\includegraphics[width=0.8\textwidth]{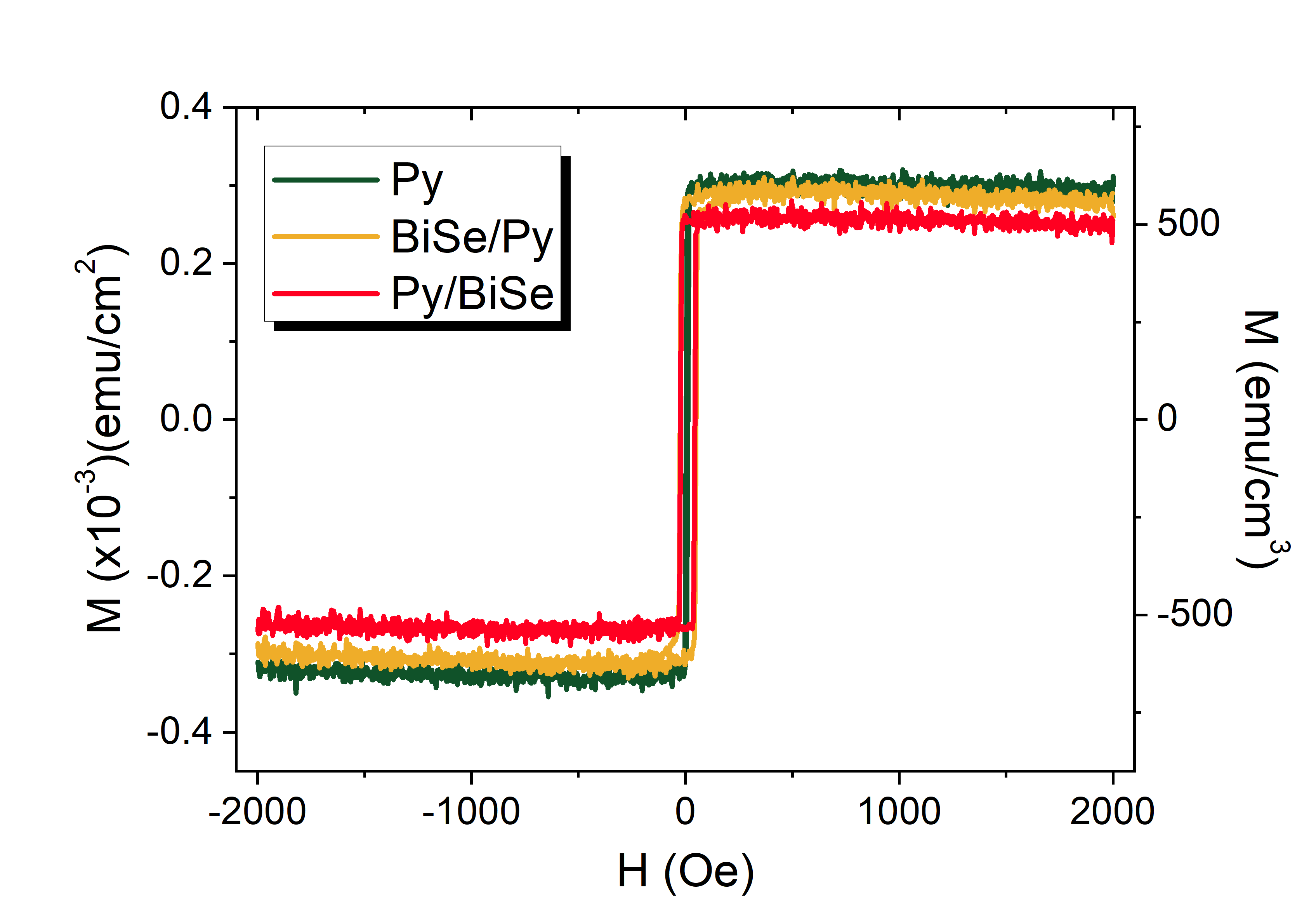}
\caption{\label{fig:vsm} \textbf{Magnetometry measurements of the Py/BiSe, BiSe/Py and Py/Pt} bilayers for a different set of samples grown under the same conditions as the ones in the main part of the manuscript. The hysteresis loops were measured at room temperature with a vibrating sample magnetometer (VSM). The magnetization by unit volume in emu/cm$^3$ is calculated considering a Py thickness of 5 nm, while the moment per unit area is calculated normalizing by the surface of the sample measured in VSM.}
\end{figure}

Figure \ref{fig:vsm} shows the in-plane hysteresis loops obtained by VSM for the new set of samples. We observe the same trend for the saturation magnetization (\textit{M}\textsubscript{s}) as in the measurements of the effective magnetization (\textit{M}\textsubscript{eff}) reported in the main text, but the absolute values differ, especially for the Py/BiSe sample, whose \textit{M}\textsubscript{s} here is around 500 emu/cm$^3$ compared to the 316 obtained for \textit{M}\textsubscript{eff} in the spin pumping measurements. This could be explained considering that, since there is a significant interdiffusion, and even Ni accumulated at the other side of the interface, the volume we need to calculate \textit{M}\textsubscript{s} is not well defined. In that sense, we have included in figure \ref{fig:vsm} the magnetization considering a Py layer of 5 nm (right axis), but also the magnetic moment normalized by the surface of the sample (left axis). This could be explained by changes in the growth conditions and/or deviations in the estimation of \textit{M}\textsubscript{s} due to the much lower precision of the VSM technique.

\FloatBarrier
\section{Details on the lock-in measurements}
\FloatBarrier

In our SP-FMR experiments, we modulated the amplitude of the RF signal rather than the magnetic field, and we always recorded both the real and imaginary parts of the measured voltage simultaneously. We always fix the phase to that no extra phase was added by the modulation signal. Transport effects are fast enough to appear without delay in the measurement while thermal effects that might be slower would appear in the imaginary part of the signal. To further assure this, we always recorded both the real and imaginary parts of the voltage. 

\begin{figure}[thb]
\centering
\includegraphics[width=0.5\textwidth]{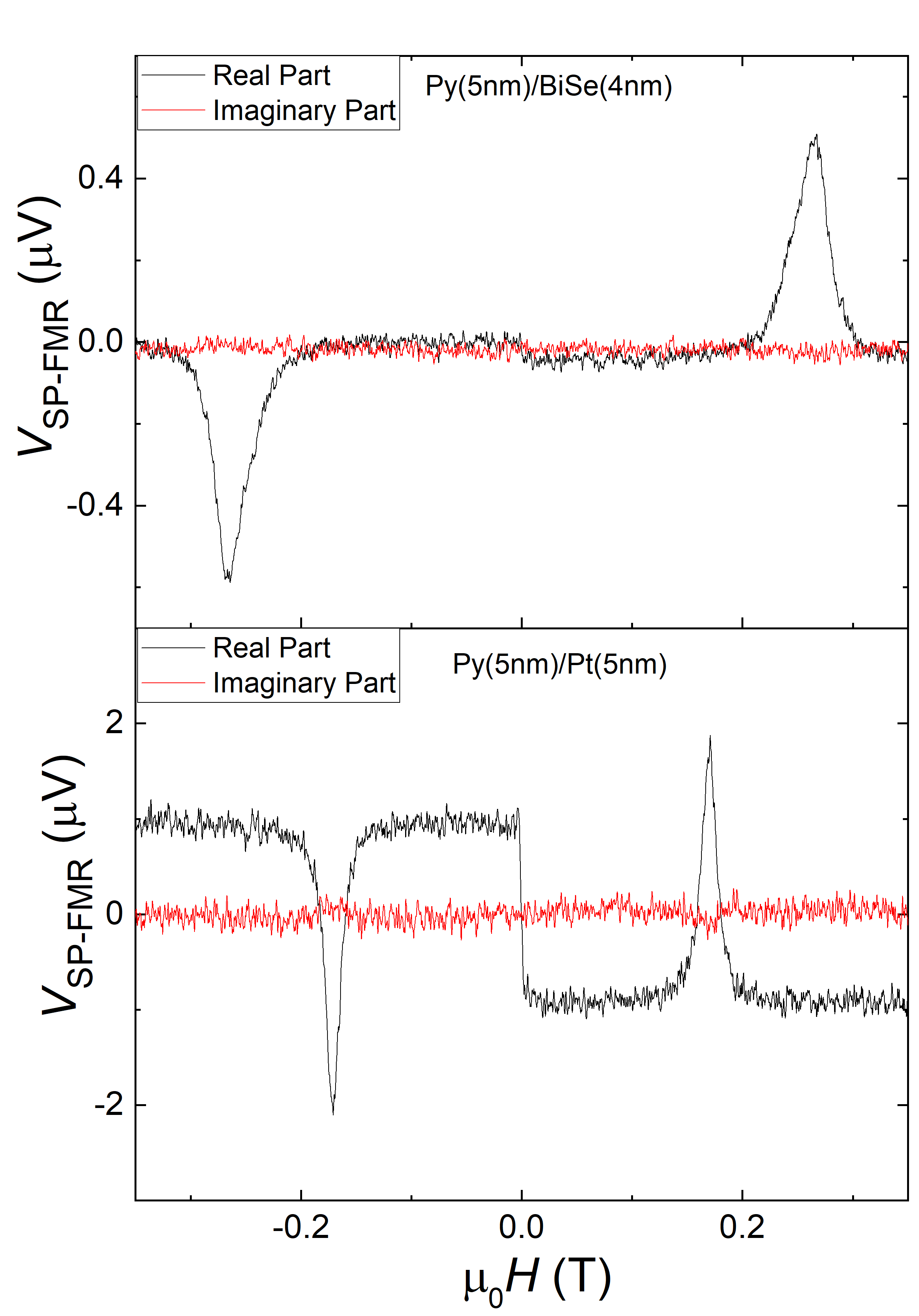}
\caption{\label{fig:RealIma} \textbf{Real and imaginary parts of the lock-in voltage detected in SP-FMR experiments} for both Py/BiSe and Py/Pt at 12GHz and 12 dBm}
\end{figure}

As an example, in Figure \ref{fig:RealIma}, we show the real and imaginary parts of the Lock-in voltage detected in SP-FMR experiments for both Py/BiSe and Py/Pt samples. The voltage signal always appears in the real part of the lock-in measurement. Repeated measurements with nominally the same layer structure on multiple devices lead to deviations in the spin pumping voltage below 5\%. 

\FloatBarrier

\bibliographystyle{apsrev4-2}
\bibliography{Bibliography.bib}

\end{document}